\documentclass[twocolumn,aps,prb,showpacs,amsmath,amssymb]{revtex4-1}
\usepackage{epsfig}
\usepackage{graphicx}
\usepackage{dcolumn}
\usepackage{bm}
\usepackage[colorlinks=true,dvipdfm]{hyperref}
\usepackage{multirow}

\begin{document}
\title{Fermion-induced quantum critical point in the Landau-Devonshire model}
\author{Shuai Yin}
\email{sysuyinshuai@gmail.com}
\author{Zhi-Yao Zuo}
\affiliation{School of Physics, Sun Yat-Sen University, Guangzhou 510275, P. R. China}
\date{\today}

\begin{abstract}
Fluctuations can change the phase transition properties drastically. An example is the fermion-induced quantum critical point (FIQCP), in which fluctuations of the massless Dirac fermions turn a putative Landau-de Gennes first-order phase transition (FOPT) with a cubic boson interaction into a continuous one. However, for the Landau-Devonshire theory, which characterizes another very large class of FOPTs, its fate under the coupling with extra fluctuations has not been explored. Here, we discover a new type of FIQCP, in which the Dirac fermion fluctuations round the boson Landau-Devonshire FOPT into a continuous phase transition. By using the functional renormalization group analyses, we determine the condition for the appearance of this FIQCP. Moreover, we point out that the present FIQCP can be a supersymmetric critical point. We finally show that the low-temperature phase diagram can provide distinct experimental evidences to detect this FIQCP.

\end{abstract}
\maketitle
\section{Introduction}
Understanding universal behaviors in phase transitions is one of the central and challenging issues in modern condensed matter physics and statistical mechanics~\cite{Sachdevbook,Wen,Fradkin}. Landau and Ginzburg proposed that the free energy can be expanded as a series of the order parameter~\cite{Landau}. According to the divergences of the derivation of the free energy, phase transitions can be classified into first-order phase transitions (FOPTs) and continuous ones~\cite{Landau,Fisher1967}. Generally, at the transition point of an FOPT, there is an energy barrier between two phases and the order parameter jumps discontinuously via nucleation and growth. According to the Landau-Ginzburg theory, there are various typical FOPT models~\cite{Landau}. The first example is the phase transition between two ordered phases~\cite{Landau,Fisher1967} and the second one is the Landau-de Gennes model with a cubic term in its free energy~\cite{Landau,Fisher1967}. Besides, the Landau-Devonshire model with a negative quartic term in its free energy~\cite{Landau,Devonshire} describes another kind of FOPTs in plenty of lattice models and real materials, ranging from ferroelectric phase transitions~\cite{Toledano}, magnetic phase transitions~\cite{Blume,Capel}, to non-fermi liquids in itinerant fermionic systems~\cite{Jakubczyk,Jakubczyk2009}.
%


To what extent fluctuations can modify the picture of the FOPT is ambiguous. Except for the $(1+1)$-dimensional (D) Potts model, usually it is believed that fluctuations of the order parameter cannot change FOPT qualitatively due to the presence of the energy barrier at the transition point~\cite{Gunton1983}. However, recently this paradigm has been challenged by various examples, in which the FOPTs can be rounded into continuous ones by the fluctuations from other degrees of freedom. For the first kind of FOPT, the theory of deconfined quantum critical point (DQCP)~\cite{Senthilsci2004,Senthilprb2004,Sandvik2007,Nogueira2007,Melko2008,Block2013,Lou2009,Pujari2013,Nahum2015A,Shao2016,Nahum2015B,Sato2017,Sreejith2019} demonstrates that a continuous phase transition happens between two ordered phases as a result of the fluctuations of emergent spinons and gauge fields~\cite{Senthilsci2004,Senthilprb2004}. A prominent example is the phase transition between the Neel order and the valence bond solid (VBS) in the $(2+1)$D Heisenberg model~\cite{Senthilsci2004,Senthilprb2004}. For the second kind of the FOPT, the theory of fermion-induced quantum critical point (FIQCP)~\cite{Li2017,Scherer2016,Classen2017,Jian2017A,Jian2017B,Torres2018,LiB2019}, which will be referred to as type-I FIQCP, shows that the putative Landau-de Gennes FOPT is turned into a continuous one by the fluctuations from the Dirac fermions. An evidence for the type-I FIQCP has been discovered in the phase transition from the Dirac semimetal to the Kekule-VBS in $(2+1)$D honeycomb lattice~\cite{Li2017}. Both cases have stimulated enormous interests in theoretical, numerical and experimental aspects~\cite{Senthilsci2004,Senthilprb2004,Sandvik2007,Nogueira2007,Melko2008,Block2013,Lou2009,Pujari2013,Nahum2015A,Shao2016,Nahum2015B,Sato2017,Sreejith2019,Li2017,Scherer2016,Classen2017,Jian2017A,Jian2017B,Torres2018,Roy,LiB2019,Zhou2016}. However, how the extra fluctuations modify the Landau-Devonshire FOPT has not been explored.




On the other hand, investigations on critical behaviors in correlated Dirac systems have attracted persistent attentions~\cite{Herbut2006,Honerkamp2008,Herbut2009,Strack2010,Yao2015,Mihaila2017,Zerf2017,Ihrig2018,Meng2019,Lang2019,herbutlor,Mengzy} as a result of their theoretic importance in reflecting some mysteries in high-energy physics and potential applications in electron devices~\cite{Geim2009RMP,Kane2010RMP,Zhang2011RMP}. Despite the simple linear dispersion, the Dirac systems provide very rich phase diagrams and colorful critical phenomena with different forms of interactions. For example, the supersymmetry, which was originally introduced in particle physics to attempt to solve various fundamental issues such as the hierarchy problem~\cite{Wess1974,Weinberg}, have been discovered as an emergent critical phenomenon in some topological Dirac systems~\cite{ssLee2007,Vishwanath2014}.
\begin{figure}
  \centering
   \includegraphics[bb= 0 0 550 595, clip, scale=0.35]{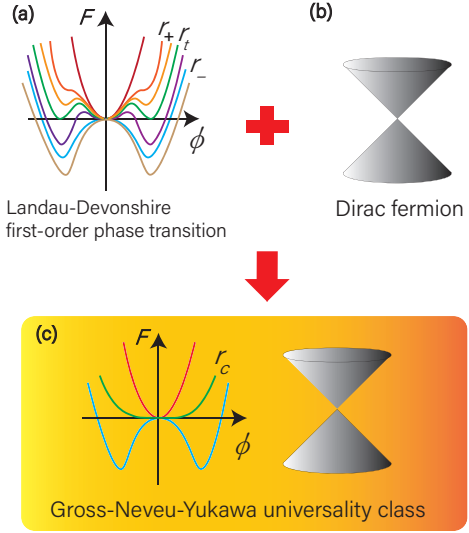}
   \caption{\label{type2} Schematic diagram of the type-II FIQCP. The Landau-Devonshire FOPT (a) with a negative quartic coupling is converted into a continuous phase transition (c) by the fluctuations of the Dirac fermions (b). The renormalization group shows that the resulting continuous phase transition belongs to the GNY universality class. In (a), $r_{\rm +}$ and $r_{\rm -}$ denotes the spinodal decomposition points for the ordered phase and the disordered phase, respectively, and $r_t$ denotes the phase transition point. In (c), $r_c$ is the critical point of FIQCP.}
\end{figure}

Inspired by these, here we explore the fate of the Landau-Devonshire FOPT in the presence of external fluctuations provided by the Dirac fermions. We discover a new type of FIQCP, which will be termed as type-II FIQCP in the following, by showing that the fluctuations of the Dirac fermions can round the Landau-Devonshire FOPT into a continuous one, as illustrated in Fig.~\ref{type2}. We find that the negative coefficient of the quartic boson coupling at the ultraviolet (UV) scale can flow to a positive value at the infrared (IR) low-energy scale with the aid of the fermion fluctuations. As a result, the energy barrier at the FOPT disappears and the order parameter changes continuously at the phase transition. Moreover, we determine the condition for the appearance of the type-II FIQCP and a trictitical point is found. We then show that the universal behavior near this type-II FIQCP is characterized by the Gross-Neveu-Yukawa (GNY) universality class~\cite{Gross1974,Rosenstein1993}, similar to the type-I FIQCP~\cite{Li2017}. However, we show that the type-II FIQCP demonstrates qualitatively different critical properties compared with the type-I case. In particular, we show that the supersymmetric criticality is permitted in the type-II FIQCP, but it has been ruled out in the type-I case~\cite{Li2017,Classen2017}. We also point out that the thermal phase diagram can provide sharp evidences in experiments to detect the FIQCP. The lattice model is also discussed. This makes the experimental examination more feasible. Since both first-order and continuous phase transitions are involved, we resort to the non-perturbative functional renormalization group (FRG) method~\cite{Wetterich1993,Berges2002}, which has been proved to be a very powerful tool in investigations on both classical and quantum phase transitions~\cite{Janssen2014B,Scherer2016,Classen2017,Gracey2018,Knorr2018}.

The rest of the paper is organized as follows. After introducing the effective action in Sec.~\ref{action}, we show the FRG analyses for the type-II FIQCP in Sec.~\ref{frgresults}. In Sec.~\ref{thermal}, we further study the thermal phase diagram near the type-II FIQCP. Then we discuss some relevant lattice models in Sec.~\ref{Discussion}. Finally, a summary is given in Sec.~\ref{Summary}.

\section{Action}\label{action}
The action of the system includes three parts. The first part is the pure bosonic Landau-Devonshire model,
\begin{eqnarray}
\begin{aligned}
S_\phi=\int d^Dx \left[\frac{1}{2}(\partial_\mu\tilde{\phi}_a)^2+\tilde{u}(\tilde{\rho})\right],
\label{eaction1}
\end{aligned}
\end{eqnarray}
in which $D$ is the space-time dimension, $\mu$ runs from $0$ to $(D-1)$, $\tilde{\phi}$ is the order parameter field, $a$ denotes the component of the bosonic field and runs from $1$ to $S$, and $\tilde{u}(\tilde{\rho})$ is the boson potential, which reads
\begin{equation}
\tilde{u}(\tilde{\rho})=\sum_{n=1}^{\infty}\tilde{\lambda}_{2n}\tilde{\rho}^n,
\label{bpotential}
\end{equation}
with $\tilde{\rho}\equiv\sum_a(\tilde{\phi}^2_a)/2$, $\tilde{\lambda}_2=r$ and $\tilde{\lambda}_4<0$. As a result of this negative quartic coupling, the Landau-Devonshire model describes a kind of FOPT~\cite{Landau,Devonshire,Toledano}. At its transition point $r_t$, the free energies are equal for the ordered phase and disordered phase, as shown in Fig. \ref{type2} (a). Between them, there is a energy barrier proportional to $-\tilde{\lambda}_4^3$ and the order parameter jumps discontinuously to cross this barrier via nucleation and growth. The second part is the Dirac fermion action, which is given by
\begin{eqnarray}
\begin{aligned}
S_\psi=&\int d^Dx [\tilde{\bar{\psi}}(\mathcal{I}_2\otimes\gamma_\mu)\partial_\mu\tilde{\psi}],
\label{eaction2}
\end{aligned}
\end{eqnarray}
in which $\mathcal{I}_2$ means $2$-dimensional identities, and $\gamma_\mu$ are ($4\times4$) matrices satisfying the Clifford algebra, i.e., $\{\gamma_\mu,\gamma_\nu\}=2\delta_{\mu\nu}$. The third part is the coupling between fermionic and bosonic fields,
\begin{eqnarray}
\begin{aligned}
S_{\psi\phi}=&\int d^Dx [\tilde{g}\tilde{\phi}_a\tilde{\bar{\psi}}(\sigma_a\otimes\mathcal{I}_4)\tilde{\psi}],
\label{eaction3}
\end{aligned}
\end{eqnarray}
in which $\tilde{g}$ is the Yukawa coupling, and $\sigma_a$ is the Pauli matrix in the $a$th direction. Although the fermionic flavor number is chosen to be $N_f=2$ (for the spinful electron in honeycomb lattice) in Eqs.~(\ref{eaction2}-\ref{eaction3}), we will also consider general cases for arbitrary $N_f$.

\section{FRG Analyses}\label{frgresults}
\subsection{RG flow equation}
The FRG method begins with the Wetterich equation~\cite{Berges2002,Wetterich1993}
\begin{equation}
\frac{\partial \Gamma_k}{\partial t}=\frac{1}{2}{\rm STr}\left[(\Gamma_k^{(2)}+R_k)^{-1}\frac{\partial R_k}{\partial t}\right],
\label{wetterich}
\end{equation}
in which $t\equiv{\rm ln} \frac{k}{\Lambda}$ with $\Lambda$ being the UV scale and $k$ being the running momentum scale, $\Gamma_k$ is the effective average action, satisfying $\Gamma_{k\rightarrow\Lambda}= S$ and $\Gamma_{k\rightarrow0}= \Gamma$ with $\Gamma$ being the effective action, $R_k$ is the regulator function, ${\rm STr}$ denotes the supertrace, and $\Gamma_k^{(2)}(i,j)\equiv\frac{\overrightarrow{\delta}}{\delta\Phi_i}\Gamma_k\frac{\overleftarrow{\delta}}{\delta\Phi_j}$
with $\Phi\equiv(\phi_a,\bar{\psi},\psi)$. In addition, the dimensionless variables, defined as
\begin{eqnarray}
\rho &\equiv& Z_{\phi,k} k^{2-D} \tilde{\rho}, \quad \quad \psi\equiv \sqrt{Z_{\psi,k}} k^{1-D} \tilde{\psi}, \\
g^2 &\equiv & k^{D-4}Z_{\phi,k}^{-1}Z_{\psi,k}^{-2}\tilde{g}^2, \quad u(\rho) \equiv k^{-D} U({\tilde\rho}),
\end{eqnarray}
are needed to explore the fixed point information.


To solve the Wetterich equation~(\ref{wetterich}), we employ the modified local potential approximation (LPA')~\cite{Berges2002} with the Litim regulators~\cite{Litim2001}, i.e., $R_{k}(q)=Z_{\phi,k}(k^2-q^2)\Theta(k^2-q^2)$ for bosonic fields and $R_{k}(q)=Z_{\Psi,k}iq_\mu\gamma_\mu(k/q-1)\Theta(k^2-q^2)$ for fermion fields, in which $\Theta$ is the step function. In this way, one obtains the RG equations for the dimensionless boson potential and Yukawa coupling as follows,
\begin{eqnarray}
\partial_t u(\rho) =&& -Du(\rho)+(D-2+\eta_{b,k})\rho u'(\rho) \nonumber\\
&&~~ +2(S-1)v_D l_0^{(B)}(u'(\rho),\eta_{b,k}) \nonumber\\
&&~~ +2v_D l_0^{(B)}(u'(\rho)+2\rho u''(\rho),\eta_{b,k}) \nonumber\\
&&~~ -2v_DN_fd_\gamma l_0^{(F)}(2h^2\rho,\eta_{f,k}), \label{beta1}\\
\partial_t g^2 =&& (D-4+\eta_{b,k}+2 \eta_{f,k})g^2 \nonumber\\
&&~~-8(S-2)v_Dl_{11}^{(FB)}(u'(0),\eta_{b,k},\eta_{f,k}), \label{beta4}
\end{eqnarray}
in which $v_D^{-1}\equiv2^{D+1}\pi^{D/2}\Gamma(D/2)$, all $l$s are threshold functions (see Appendix~\ref{appendix1}), the minus sign before the last term in Eq.~(\ref{beta1}) is the sign factor from the fermion loop, and the running anomalous dimensions for boson and fermion fields, defined as $\eta_{b,k}\equiv-\partial_t Z_{\phi,k}/Z_{\phi,k}$ and $\eta_{f,k}\equiv-\partial_t Z_{\psi,k}/Z_{\psi,k}$, with $Z_{\phi/\psi}$ being field-renormalization factors for boson/fermion fields, are solved as
\begin{eqnarray}
\eta_{b,k} &=&\frac{4 v_D}{D}2 N_f d_\gamma g^2 m_4^{(F)}(\eta_{f,k}), \label{eta1} \\
\eta_{f,k} &=& \frac{8 v_D}{D} S g^2 m_{12}^{(FB)}(u'(0),\eta_{b,k}),
\label{eta2}
\end{eqnarray}
in which $m$s are also threshold functions (see Appendix~\ref{appendix1}).

\subsection{Mechanism of the type-II FIQCP}
Before the full FRG calculation, let us at first focus on the RG flow of $\lambda_4$ and give a heuristic analysis. Since the energy barrier at the FOPT of the Landau-Devonshire model is induced by the negative $\lambda_4$, the possibility of the FIQCP is reflected in the impact of the coupling with fermions on $\lambda_4$. By deriving Eq.~(\ref{beta1}) with respect to $\rho$, one obtains the RG equation for $\lambda_4$,
\begin{equation}
\frac{d\lambda_4}{dt}=\lambda_4-\frac{6}{\pi^2}\lambda_4^2+\frac{8N_f g^4}{3 \pi^2},
\label{lambda4}
\end{equation}
in which $S=1$ is chosen as an example and only the leading terms are kept. In the right hand side of Eq.~(\ref{lambda4}), the first term comes from the engineering dimension, the second term comes from the boson fluctuation, while the third term comes from the Dirac fermion fluctuation. Without coupling to fermions, i.e., $g=0$, Eq.~(\ref{lambda4}) shows that a negative $\lambda_4$ in the UV scale will decrease monotonously without a lower bound as the energy scale goes down along the IR direction. This corresponds to the Landau-Devonshire FOPT. But, remarkably, this situation can be changed when the coupling to the fermion is introduced. The anti-commutativity of the fermion fields leads to an additional minus sign factor in the fermionic loop diagram. This makes the last term in Eq.~(\ref{lambda4}) positive. Accordingly, the last term makes the opposite contribution in contrast to the previous two terms. Furthermore, when $g$ is large enough, the right hand side of Eq.~(\ref{lambda4}) can be positive. Thus, this fermion term can propel $\lambda_4$ to flow to positive values.

On the other hand, in the boson part, whether there is an FOPT is determined by the sign of $\lambda_4$. A negative $\lambda_4$ leads to an FOPT, while a positive $\lambda_4$ corresponds to a continuous phase transition. Concretely, we approximate the model by neglecting the terms higher than $\lambda_6 \rho^3$. The variational solution of the order parameter at the maximum value of the energy barrier of the FOPT is~\cite{Devonshire,Toledano}
\begin{equation}
\langle\phi\rangle_{\rm max}=\pm\sqrt{\frac{-\lambda_4-\sqrt{\lambda_4^2-3\lambda_2\lambda_6}}{6\lambda_6}}.
\label{maxop}
\end{equation}
From Eq.~(\ref{maxop}), one finds that an unphysical imaginary part will be developed when $\lambda_4$ becomes positive. Accordingly, a positive $\lambda_4$ signifies the disappearance of the energy barrier of the FOPT. Since this change of sign in $\lambda_4$ is induced by the fermion fluctuations, an FIQCP can then arise. Although the argument above combines the RG flow Eq.~(\ref{lambda4}) with the variational solution and is not an exact one, it still reveals the main mechanism of the type-II FIQCP.

\begin{figure}
  \centerline{\epsfig{file=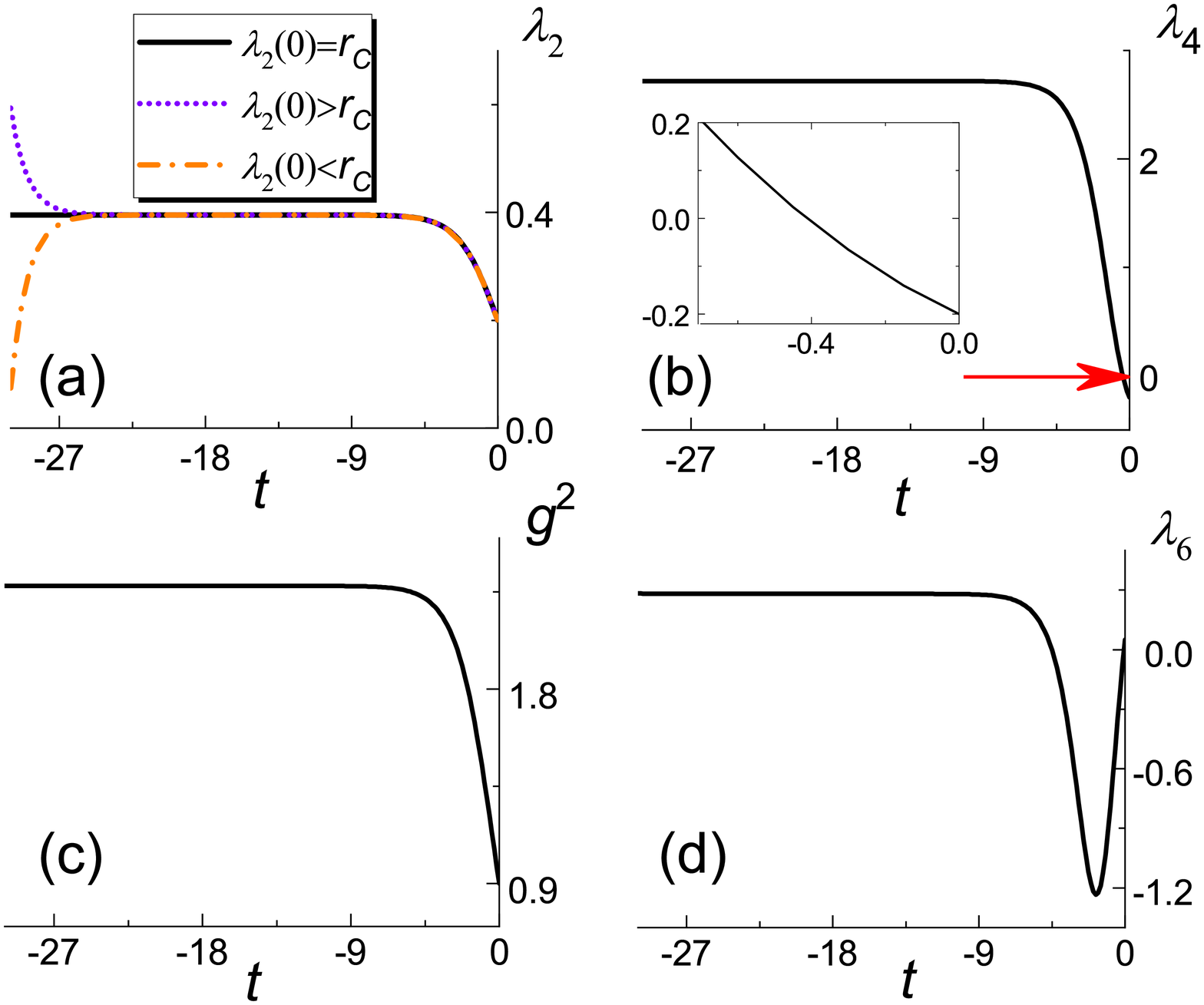,width=0.9\columnwidth}}
  \caption{\label{rgflowlam} The RG flows for the type-II FIQCP. With $N_f=2$ and $S=1$, the RG flows of different parameters running from $t=0$ (UV) to $t\rightarrow-\infty$ (IR) are shown in (a-d). By tuning $\lambda_2$ at UV scale, an IR fixed point is found at $\lambda_2(0)=r_{C}$. (a) shows the RG flow for $\lambda_2$ at and near the critical point. (b-d) show the RG flows at $\lambda_2=r_C$ for $\lambda_4$, $g^2$, and $\lambda_6$, respectively. In particular, $\lambda_4$ runs from a negative value to a positive one, as shown in the inset of (b). The arrow in (b) denotes the zero point. The bare parameters are chosen as $\lambda_4=-0.2$, $g^2=0.9$, $\lambda_6=0.05$, $\lambda_8=\lambda_{10}=\lambda_{12}=0.1$ and $r_C\simeq0.199727$.
  }
\end{figure}
\subsection{Fixed point and universality class}
To quantitatively investigate the critical properties of this type-II FIQCP, we solve the RG equations~(\ref{beta1}) and~(\ref{beta4}) numerically. We will take the case of $S=1$ and $N_f=2$ as an example and truncate Eq.~(\ref{bpotential}) to the sixth order. Figure~\ref{rgflowlam} shows the RG flow for a set of arbitrary bare parameters with $\lambda_4<0$ and $g$ being large enough. We have checked that the pure bosonic Landau-Devonshire theory with the same set of initial parameters does not possess any finite infrared fixed point. However, form Fig.~\ref{rgflowlam}, one finds that the RG flow runs towards a fixed point when $\lambda_2(0)$ is finely tuned. This means $\lambda_2$ is the only relevant scaling variable. According to the RG theory of the critical phenomena~\cite{Wilson1974}, a continuous phase transition is dictated by a single relevant variable when the symmetry-breaking field is absent. Therefore, the RG flows in Fig.~\ref{rgflowlam} (a) shows that a continuous phase transition arises with $\lambda_2$ being its relevant direction. Given that the pure boson part features an FOPT, one concludes that this continuous phase transition is just the FIQCP. Moreover, Fig.~\ref{rgflowlam} (b) shows that $\lambda_4$ changes its sign from negative to positive at some intermediate scale, consistent with the discussion above. Also, Fig.~\ref{rgflowlam} (c) shows that $g^2$ tends to a nonzero value, demonstrating that the interaction between the Dirac fermion and the boson plays indispensable roles. In addition, we also identify the type-II FIQCP for other cases with different $S$ and $N_f$ (see Appendix~\ref{appendix2}).

We then calculate the critical exponents. The boson and fermion anomalous dimensions are determined from Eqs.~(\ref{eta1}) and (\ref{eta2}), and the correlation length exponent $\nu$ is estimated from the positive eigenvalue of the stability matrix, which is defined as $M_{i,j}\equiv-\partial B_i/\partial \alpha_j$ with $\alpha_i\equiv(\lambda_{2n}, g^2)$ and $B_i\equiv(\partial_t\lambda_{2n},\partial_t g^2)$. We list the exponents in Table.~\ref{tabexp}. By comparing these exponents with the known results~\cite{Ihrig2018,Mihaila2017,Zerf2017,Janssen2014B,Gracey2018,Knorr2018}, we find that the type-II FIQCP belongs to the GNY universality class.
\begin{table}
  \centering
  \caption{Critical exponents for the type-II FIQCP.}
    \begin{tabular}{c p{1.2cm} p{1.2cm} p{1.2cm}}
    \hline
    \hline
    \multicolumn{1}{c|}{}~~~ &$1/\nu$ &$\eta_b$ & $\eta_f$  \\
    \hline
    \multicolumn{1}{c|}{$N_f=2$, $S=1$}{}~~~ &0.981   &0.759 & 0.032  \\
    \multicolumn{1}{c|}{$N_f=2$, $S=2$}{}~~~ &0.862   &0.875 & 0.062  \\
    \multicolumn{1}{c|}{$N_f=2$, $S=3$}{}~~~ &0.771   &1.015 & 0.084  \\
    \multicolumn{1}{c|}{$N_f=1/2$, $S=2$}{}~~~ &1.062 &0.353 & 0.323   \\
    \hline
    \hline
    \end{tabular}%
  \label{tabexp}%
\end{table}%

\subsection{Chiral tricritical point}
\begin{figure}
  \centerline{\epsfig{file=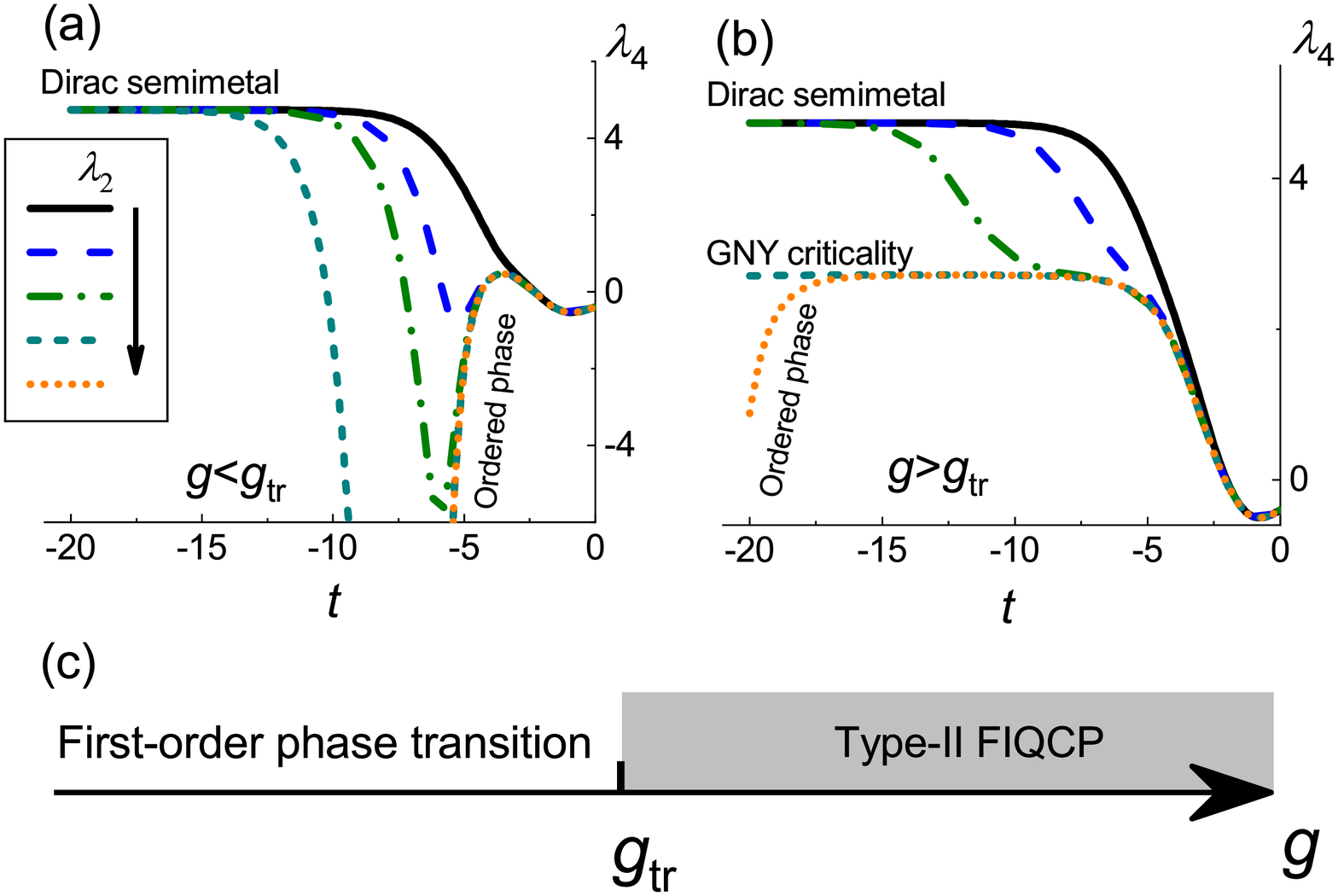,width=0.9\columnwidth}}
  \caption{\label{rgflowgtr} A tricritical point $g_{\rm tr}$ is discovered from the RG flows of $\lambda_4$ for different $g$. When $g>g_{\rm tr}$, by tuning $\lambda_2$, one finds that besides the Dirac semimetal phase and the ordered phase, the GNY fixed point can also be found as shown in (b), in contrast, when $g<g_{\rm tr}$, there is no such critical fixed point (a). This shows that the type-II FIQCP can only appear for $g>g_{\rm tr}$ (c). The bare parameters are chosen as $\lambda_4=-0.4$, $\lambda_6=0.05$, and $\lambda_8=\lambda_{10}=\lambda_{12}=0.1$. For these parameters, $g_{\rm tr}$ is estimated as $g^2_{\rm tr}\simeq 0.47$. $g^2$ is chosen as $g^2=0.4$ and $g^2=0.48$ in (a) and (b), respectively. The legend shows different $\lambda_2(0)$ sorted in descending order. Note that in the semimetal phase, although $\lambda_4$ tends to a finite value in the IR limit~\cite{Torres2018}, $\lambda_2$ diverges as shown in Fig.~\ref{rgflowlam}.
  }
\end{figure}

In this section, we determine the condition for the appearance of the type-II FIQCP. Figure~\ref{rgflowgtr} shows that for large $g$, besides the Dirac semimetal phase with a finite $\lambda_4$~\cite{Torres2018} in the IR limit and the ordered phase with an infinite $\lambda_4$ in the IR limit, the GNY fixed point appears by tuning $\lambda_2$, however, for small $g$ this GNY fixed point is absent. This demonstrates that the type-II FIQCP arises only for large enough $g$. Intuitively, for small $g$, the coupling between fermionic and bosonic modes is too weak to soften the FOPT. Thus, there is a tricritical point, $g_{\rm tr}$, only when $g>g_{\rm tr}$ can the FIQCP arise. In addition, from Eq.~(\ref{lambda4}), we conclude that $g^4_{\rm tr}$ increases as the absolute value of $\lambda_4$ increases and is inversely proportional to $N_f$. These deductions are verified in Appendix~\ref{appendix3}. Furthermore, by noting the essential roles played by the Dirac fermions, we conclude that this tricritical point is just the chiral tricritical point~\cite{Yin2018}.

The appearance of this chiral tricritical point demonstrates that the mechanism of the type-II FIQCP is quite different from that of the type-I case~\cite{Li2017}, since there is no such tricrtical point in the latter case. Instead, the condition for the appearance of the type-I FIQCP is that the fermion flavor number $N_f$ should be large enough. The reason is that the type-I FIQCP is completely governed by the dimension of the cubic term at the GNY fixed point. For small $N_f$, the cubic term is relevant since its scaling dimension near the GNY fixed point is positive. In contrast, the type-II FIQCP is mainly determined by bare parameters. For a UV negative $\lambda_4$, the phase space spanned by bare parameters is divided into two regions as shown in Fig.~\ref{rgflowgtr} (c): one corresponds to the FOPT without any finite fixed point, the other corresponds to the FIQCP with $\lambda_2$ being its only relevant direction, while in between there is a chiral tricritical point as discussed above. As shown in Ref.~\cite{Yin2018}, this chiral tricritical point has two relevant directions: one is $\lambda_2$, the other lies in the plane spanned by $g$ and $\lambda_4$. Since the physical picture discussed in Fig.~\ref{rgflowgtr} (c) is independent of $N_f$, the type-II FIQCP can appear for any $N_f$.

\subsection{Supersymmetric critical point}
\begin{figure}[tbp]
  \centerline{\epsfig{file=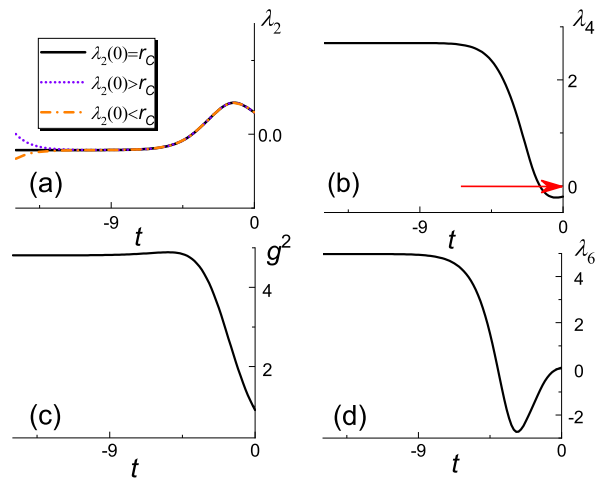,width=0.9\columnwidth}}
  \caption{\label{rgflowsusy} For $N_f=1/2$ and $S=2$, the RG flows running from $t=0$ (UV) to $t\rightarrow-\infty$ (IR) are shown in (a-d). The bare parameters are chosen as $\lambda_4(0)=-0.2$, $g^2(0)=0.9$, $\lambda_6(0)=0.05$, $\lambda_8(0)=\lambda_{10}(0)=\lambda_{12}(0)=0.1$. By tuning $\lambda_2$, a fixed point can be found at $\lambda_2(0)=r_{C}=0.05919299198090491$. (a) shows the RG flows for $\lambda_2$ at and near the critical point, as marked. (b-d) show the RG flow at $\lambda_2(0)=r_C$ for $\lambda_4$, $g^2$, and $\lambda_6$, respectively. In particular, $\lambda_4$ runs from a negative value to a positive one as shown in (b). The arrow in (b) denotes the zero point.
  }
\end{figure}

For the case of $N_f=1/2$ and $S=2$, the GNY universality class corresponds to an emergent supersymmetric fixed point described by the Wess-Zumino theory~\cite{Wess1974}. Scaling analyses show the type-I FIQCP cannot occur in these systems, since the cubic term is always relevant near this supersymmetric fixed point~\cite{Li2017,Scherer2016,Classen2017} and it drives the RG flow to an infinite fixed point. Accordingly, when $N_f=1/2$ and $S=2$ the Dirac system with the Landau-de Gennes type bosonic interaction exhibits an FOPT. In contrast, for the Landau-Devonshire bosonic model, we show in Fig.~\ref{rgflowsusy} that a continuous phase transition arises by tuning $\lambda_2$ for large enough $g$. The supersymmetric property requires that $\lambda_4=g^2/2$~\cite{ssLee2007,Vishwanath2014} in the IR limit. In Fig.~\ref{rgflowsusy}, $\lambda_4 \simeq2.693$ and $g^2\simeq4.808$ at the IR fixed point. These results agree to the supersymmetric condition within $12\%$. The deviation may come from the cutoff scheme and some improved procedures have been proposed in recent studies~\cite{susyimprove}.

Moreover, the anomalous dimensions for the fermion and boson fields near the supersymmetric critical point for $N_f=1/2$ and $S=2$ can be exactly solved as $\eta_\psi=\eta_\phi=1/3$~\cite{Aharony1997}. To further check whether the fixed point found in Fig.~\ref{rgflowsusy} is the supersymmetric critical point, we calculate the critical exponents as shown in Table~\ref{tabexp}. These results agree within $6\%$ to the exact results, demonstrating the appearance of the supersymmetric critical point~\cite{Classen2017}. Also, it is expected that more efficient cutoff scheme can give more accurate results~\cite{susyimprove}.

\section{Thermal phase diagram} \label{thermal}
\begin{figure}[t]
  \centering
   \includegraphics[bb= 0 0 360 310, clip, scale=0.42]{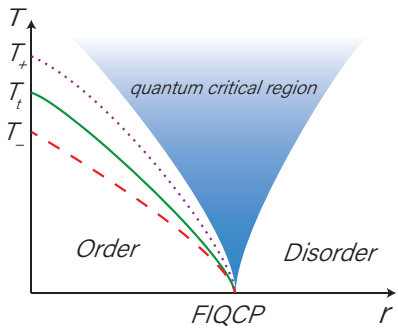}
   \caption{\label{qcregion} Schematic phase diagram near an FIQCP. The fan of the quantum critical region is dominated by the zero-temperature FIQCP. Besides, we show the classical phase transition is an FOPT, whose transition point is denoted by $T_t$ (solid). $T_+$ (dotted) and $T_-$ (dashed) are the spinodal points for the ordered phase and disordered phase, respectively. This phase diagram can provide sharp evidence of the type-II FIQCP in experiments.}
\end{figure}

A crucial question is how to distinguish (a) the type-II FIQCP with $\lambda_4(0)<0$ from (b) usual GNY phase transition with $\lambda_4(0)>0$ in experiments. Concretely, we consider the case for $S=1$. Although both (a) and (b) share the same zero-temperature quantum critical properties, we will show that their low-temperature thermal phase transitions are quite different. In (b), there is a continuous classical phase transition at the low-temperature region. Near this classical critical point, fermionic fluctuations are completely decoupled in low frequencies as a result of the finite Matsubara-frequency-gap propotional to the temperature $T$~\cite{Stephanov1995,Hesselmann2016}. Therefore, the classical phase transition belongs to the usual Wilson-Fisher $2$D-Ising universality class. In contrast, in (a), if the corresponding finite-temperature phase transition were changed to be continuous by fermion fluctuations, the fermionic modes should also be decoupled. In the light of this reductio ad absurdum, we conclude that the thermal phase transition is an FOPT, as illustrated in Fig.~\ref{qcregion}. Since all experiments are implemented at finite temperatures, the thermal phase diagram can provide distinct evidence to detect the type-II FIQCP in experiments.

In addition, macroscopic physical quantities near the quantum critical point must obey the scaling properties controlled by the quantum critical point. At low temperatures, the thermal FOPT should be affected by the FIQCP. For example, all three characteristic temperatures, the transition temperature $T_t$, the spinodal point for ordered phase $T_+$ and the spinodal point for disordered phase $T_-$ should be proportional to $|r-r_c|^{\nu z}$, in which $z=1$ is the dynamic exponent. Moreover, the latent heat $E_L$ between two spinocal decomposition curves with a fixed $r$ should satisfy $E_L\propto |r-r_c|^{\nu z}$. Besides, although the fermion modes cannot soft the thermal FOPT to a continuous phase transition as discussed above, it is expected that they can affect the nucleation and growth dynamics. In particular, there could be some resonance effects when the energy barrier at the transition point matches the fermion frequencies $(2n+1)\pi T$ with $n$ being an integer. Detailed studies are now in process.

\section{Discussion} \label{Discussion}
Although our study begins with a field-theoretic model, the type-II FIQCP can be realized in lattice models undoubtedly. It was shown that the FOPT and the continuous phase transition have been observed in a spin model with both nearest-neighbor and next-nearest neighbor interactions~\cite{Kato2015}. The effective theory for this model is just the Landau-Devonshire theory~\cite{Kato2015}. We can consider a similar model in graphene, whose elementary excitation is the Dirac fermion with $N_f=2$ and the spin freedom is naturally coupled to the Dirac fermion. Actually, the chiral tricritical point is proposed therein~\cite{Yin2018}. So its continuous phase transition side should be the type-II FIQCP. For the case of $N_f=1/2$ and $S=2$, in which the supersymmetric type-II FIQCP appears, it can be realized in the surface of the $3$D topological insulator~\cite{ssLee2007,Vishwanath2014,Jiansusy,Lisusy2,Rahmanisusy,Maciejkosusy}. Moreover, various synthetic cold atoms platforms hosting the Dirac fermions have been realized with tunable interactions~\cite{Greif2015,Bloch2017}. It is expected that the type-II FIQCP can be observed therein.

\section{Summary} \label{Summary}
In summary, we have found a type-II FIQCP, in which fermion fluctuations round the bosonic Landau-Devonshire FOPT into a continuous phase transition by changing the negative sign of $\lambda_4$ to be positive. Accordingly, the energy barrier at the FOPT disappears and the order parameter can change continuously. The FRG analysis has shown that this continuous phase transition belongs to the GNY universality class. We have also pointed out that to realize the type-II FIQCP, the Yukawa coupling should be larger than a tricritical value. The type-II FIQCP for $N_f=1/2$ and $S=2$ has been shown to be a supersymmetric critical point, which has been ruled out in the type-I FIQCP. We have also argued that for $S=1$ the thermal phase transition near the FIQCP is an FOPT, whose scaling behavior is affected by the FIQCP. This can provide distinctly evidences of the FIQCP in experiments. The lattice models have also been discussed.

\section*{Acknowledgements}
We wish to thank F. Zhong, H. Yao, S.-K. Jian, M. M. Scherer, L. Janssen, D. Yao, Z. Yan and L. Wang for their helpful discussions. In particular, we are grateful to M. M. Scherer for his valuable comments and suggestions. We acknowledge the support by the startup grant (\#74130-18841229) in Sun Yat-Sen University.

\appendix

\section{Threshold functions}\label{appendix1}
In the main text, the explicit formula of the FRG flow equation depends on the detailed form of the regulator. Using the Litim's regulator~\cite{Litim2001}, $R_{k}(q)=Z_{\phi,k}(k^2-q^2)\Theta(k^2-q^2)$ for bosonic fields and $R_{k}(q)=Z_{\Psi,k}iq_\mu\gamma_\mu(k/q-1)\Theta(k^2-q^2)$ for fermion fields, in which $\Theta$ is the step function, one obtains the the threshold functions as follows~\cite{Classen2017}
\begin{eqnarray}
l_0^{(B)}(w,\eta_b)=&&\frac{2}{D}\left(1-\frac{\eta_b}{D+2}\right)\frac{1}{1+w}, \label{tf1}\\
l_0^{(F)}(w,\eta_b)=&&\frac{2}{D}\left(1-\frac{\eta_f}{D+1}\right)\frac{1}{1+w}, \label{tf2}\\ \nonumber
l_{11}^{(FB)}(w,\eta_b,\eta_f)=&&\frac{2}{D}\left(1-\frac{\eta_f}{D+1}\right)\frac{1}{1+w}\\
&&+\frac{2}{D}\left(1-\frac{\eta_b}{D+2}\right)\frac{1}{(1+w)^2}, \label{tf3}\\
m_4^{(F)}(\eta_f)=&&\frac{3}{4}+\frac{1-\eta_f}{2(D-2)}, \label{tf4}\\
m_{12}^{(FB)}(w,\eta_{b})=&&\left(1-\frac{\eta_b}{D+1}\right)\frac{1}{(1+w)^2}, \label{tf5}
\end{eqnarray}
in which $w$ corresponds to the boson mass $u'(\rho)$ and the fermion mass is zero in the symmetric phase.

\section{Type-II FIQCP for other $N_f$ and $S$}\label{appendix2}
\begin{figure}[t]
  \centerline{\epsfig{file=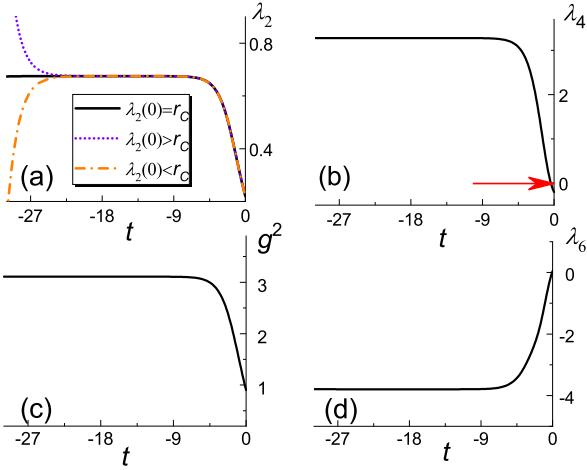,width=0.9\columnwidth}}
  \caption{\label{rgflows3} For $N_f=2$ and $S=3$, the RG flows running from $t=0$ (UV) to $t\rightarrow-\infty$ (IR) are shown in (a-d). The bare parameters are chosen as $\lambda_4(0)=-0.2$, $g^2(0)=0.9$, $\lambda_6(0)=0.05$, $\lambda_8(0)=\lambda_{10}(0)=\lambda_{12}(0)=0.1$. By tuning $\lambda_2$, a fixed point can be found at $\lambda_2(0)=r_{C}=0.2204655093005$. (a) shows the RG flow for $\lambda_2$ at and near the critical point, as marked. (b-d) show the RG flow at $\lambda_2(0)=r_C$ for $\lambda_4$, $g^2$, and $\lambda_6$, respectively. In particular, $\lambda_4$ runs from a negative value to a positive one as shown in (b). The arrow in (b) denotes the zero point.
  }
\end{figure}

In this section, we discuss the type-II FIQCP for other $N_f$ and $S$ by studying the corresponding RG flows. For $N_f=2$ and $S=3$, Fig.~\ref{rgflows3} shows the RG flows for different parameters. We find that $\lambda_2$ is the only relevant parameter, similar to the case of $S=1$. Also, we find that $\lambda_4$ also changes from a negative value to a positive fixed point. This indicates that the energy barrier at the transition point of the FOPT is smeared out by the fermion fluctuations and the Landau-Devonshire FOPT is changed to be a continuous one. The corresponding critical exponents $\nu$, $\eta_b$, and $\eta_f$ are then calculated according to the parameters at the fixed point and have been listed in Table~\ref{tabexp}. One finds that for $S=3$, the type-II FIQCP belongs to the chiral Heisenberg universality class. For the case of $N_f=2$ and $S=2$, the RG flows are quite similar. And, its critical exponents are also listed in Table~\ref{tabexp}. From their values, one finds that the corresponding type-II FIQCP belongs to the chiral XY universality class.

\section{Tricritical point for different $N_f$}\label{appendix3}

\begin{figure}[t]
  \centerline{\epsfig{file=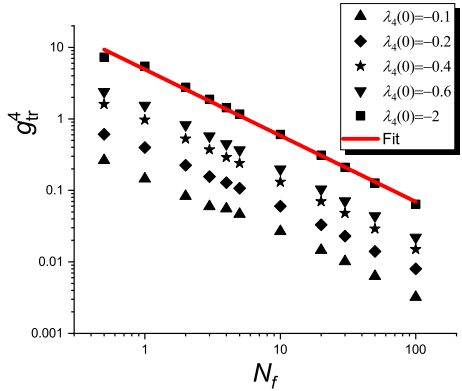,width=0.7\columnwidth}}
  \caption{\label{gtrnf} Curves of the tricritical point $g_{\rm tr}$ versus $N_f$. Other parameters are chosen as $\lambda_2(0)=0.1997273432490756$, $\lambda_6(0)=0.05$, and $\lambda_{8}(0)=\lambda_{10}(0)=\lambda_{12}(0)=0.1$. Double logarithmic scales are used. Power fitting shows that $g^4_{\rm tr}\propto 1/N_f^{0.959}$ for $\lambda_4(0)=-2$.
  }
\end{figure}

In this section, we study the dependence of the $g_{\rm tr}$ on $N_f$. We take the case of $S=1$ as an example. Directly determining $g_{\rm tr}$ from the RG flow equations is very complex as a result of the fine-tuning of $\lambda_2$. Here, we resort to the method of ``virtual" RG flow, which is constructed by multiplying a minus sign before the beta function of the $\lambda_2$ to change artificially $\lambda_2$ into an irrelevant vaiable while keeps other beta functions unchanged. This method is based on the following thoughts: With an arbitrary set of parameters, if it is controlled by a continuous phase transition with $\lambda_2$ being its only relevant direction, the resulting virtual RG flow will run towards the finite fixed point characterizing this continuous phase transition; In contrast, if it is controlled by an FOPT, in which no finite fixed point exists, the virtual flow will go towards infinity, driven by other relevant directions. Therefore, this method can be employed to classify the phase space. However, since the virtual flow is essentially different from the real one, the quantitative performance is needed to be further checked. We have checked that for the parameters chosen in Fig.~\ref{rgflowlam}, $g_{\rm tr}\simeq 0.224$ for $\lambda_2(0)=0.1997273432490756$. We also find that the minimum value of $g_{\rm tr}$ is $g_{\rm tr}\simeq 0.18$ for $\lambda_2(0)<3$. This value is close to the one obtained directly from the real flow. Also, we find that in the case of $N_f=1/2$ and $S=2$ the minimum value of $g_{\rm tr}$ is $g_{\rm tr}\simeq 0.58$ for $\lambda_2(0)<3$. This value is also close to the one obtained from the real flow.

Figure~\ref{gtrnf} shows the results for other cases. From Fig.~\ref{gtrnf} one finds that for a bare fixed $\lambda_2(0)$, the tricritical point $g_{\rm tr}$ decreases as $N_f$ increases. The reason is that for large $N_f$, fermion fluctuations are stronger and the watershed $g_{\rm tr}$ is not needed to be very large to induce a continuous phase transition. In double logarithmic scales, the curve of $g^4_{\rm tr}$ versus $N_f$ is almost a straight line. By power fitting, one finds that $g^4_{\rm tr}$ satisfies $g^4_{\rm tr}\propto1/N_f$ approximatively. Also one finds that this rule is also mainly applicable for other values of $\lambda_4(0)$. The reason can be found by inspecting Eq.~(\ref{lambda4}). From Eq.~(\ref{lambda4}), one finds that for $\lambda_4(0)<0$, the flow direction changes when $g^4(0)>\frac{3\lambda_4(0)(6\lambda_4(0)-\pi^2)}{8N_f}$. Accordingly one obtains $g^4_{\rm tr}(0)\propto1/N_f$. However, from Fig.~\ref{gtrnf}, one finds that this formula is slightly violated when the absolute value of $\lambda_4$ is small. The reason may be that other parameters, such as $\lambda_6$, can also play roles in determining $g_{\rm tr}$ for small $\lambda_4$.


\begin{thebibliography}{99}
\bibitem{Sachdevbook} S. Sachdev, \textit{Quantum Phase Transitions}(Cambridge University Press, 1999).
\bibitem{Wen}X.-G. Wen, {\it Quantum Field Theory of Many-Body Systems}, (Oxford University Press, 2004).
\bibitem{Fradkin} E. Fradkin, {\it Field Theories of Condensed Matter Physics}, 2nd ed. (Cambridge University Press, 2013).
\bibitem{Landau} L. D. Landau and E. M. Lifshitz,  {\it Statistical Physics} (Butterworth-Heinemann, 1999).
\bibitem{Fisher1967} M. E. Fisher, {\it The theory of condensation and the critical point}, Physics {\bf 3}, 255-283 (1967).
\bibitem{Devonshire} A. F. Devonshire, {\it Theory of barium titanate}, Philos. Mag. {\bf 40}, 1040-1063 (1949).
\bibitem{Toledano} J. C. Tol\'{e}dano and P. Tol\'{e}dano, {\it The Landau theory of phase transitions: Application to structural, incommensurate, magnetic and liquid crystal Systems} (World Scientific Lecture Notes in Physics, 1987).
\bibitem{Blume} M. Blume, {\it Theory of the First-Order Magnetic Phase Change in UO$_2$}, Phys. Rev. {\bf 141}, 517 (1966).
\bibitem{Capel} H. Capel, {\it On the possibility of first-order phase transitions in Ising systems of triplet ions with zero-field splitting}, Physica {\bf 32}, 966 (1966).
\bibitem{Jakubczyk} P. Jakubczyk, {\it Renormalized $\phi^6$ model for quantum phase transitions in systems of itinerant fermions}, Phys. Rev. B {\bf 79}, 125115 (2009).
\bibitem{Jakubczyk2009} P. Jakubczyk, W. Metzner, and H. Yamase, {\it Turning a First Order Quantum Phase Transition Continuous by Fluctuations: General Flow Equations and Application to d-Wave Pomeranchuk Instability}, Phys. Rev. Lett. {\bf 103}, 220602 (2009).
\bibitem{Gunton1983} J. D. Gunton, M. San Miguel, and P. S. Sahni, in {\it Phase Transitions and Critical Phenomena}, eds. C. Domb and J. L. Lebowitz Vol. 8, 267 (Academic, London, 1983).
\bibitem{Senthilsci2004} T. Senthil, A. Vishwanath, L. Balents, S. Sachdev, and M. P. A. Fisher, {\it Deconfined quantum critical points}, Science {\bf 303}, 1490–1494 (2004).
\bibitem{Senthilprb2004} T. Senthil, L. Balents, S. Sachdev, A. Vishwanath, and M. P. A. Fisher, {\it Quantum criticality beyond the Landau-Ginzburg-Wilson paradigm}, Phys. Rev. B {\bf 70}, 144407 (2004).
\bibitem{Sandvik2007} A. W. Sandvik, {\it Evidence for Deconfined Quantum Criticality in a Two-Dimensional Heisenberg Model with Four-Spin Interactions}, Phys. Rev. Lett. {\bf 98}, 227202 (2007).
\bibitem{Nogueira2007} F. S. Nogueira, S. Kragset, and A. Sudb{\o}, {\it Quantum critical scaling behavior of deconfined spinons}, Phys. Rev. B {\bf 76}, 220403(R) (2007).
\bibitem{Melko2008} R. G. Melko and R. K. Kaul, {\it Scaling in the fan of an unconventional Quantum Critical Point}, Phys. Rev. Lett. {\bf 100}, 017203 (2008).
\bibitem{Block2013} M. S. Block, R. G. Melko, and R. K. Kaul, {\it Fate of $CP^{N-1}$ fixed points with $q$ monopoles}, Phys. Rev. Lett. {\bf 111}, 137202 (2013).
\bibitem{Lou2009} J. Lou, A. W. Sandvik, and N. Kawashima, {\it Antiferromagnetic to valence-bond-solid transitions in two-dimensional SU$(N)$ Heisenberg models with multispin interactions}, Phys. Rev. B {\bf 80}, 180414(R) (2009).
\bibitem{Pujari2013} S. Pujari, K. Damle, and F. Alet, {\it Neel-State to Valence-Bond-Solid Transition on the Honeycomb Lattice: Evidence for Deconfined Criticality}, Phys. Rev. Lett. {\bf 111}, 087203 (2013).
\bibitem{Nahum2015A} A. Nahum, J. T. Chalker, P. Serna, M. Ortuno, and A. M. Somoza, {\it Deconfined Quantum Criticality, Scaling Violations, and Classical Loop Models}, Phys. Rev. X {\bf 5}, 041048 (2015).
\bibitem{Shao2016} H. Shao, W. Guo, and A. W. Sandvik, {\it Quantum criticality with two length scales}, Science {\bf 352}, 213–216 (2016).
\bibitem{Nahum2015B} A. Nahum, P. Serna, J. T. Chalker, M. Ortu\~{n}o, and A. M. Somoza, {\it Emergent SO$(5)$ Symmetry at the N\'{e}el to Valence-Bond-Solid Transition}, Phys. Rev. Lett. {\bf 115}, 267203 (2015).
\bibitem{Sato2017} T. Sato, M. Hohenadler, and F. F. Assaad, {\it Dirac Fermions with Competing Orders: Non-Landau Transition with Emergent Symmetry}, Phys. Rev. Lett. {\bf 119}, 197203 (2017).
\bibitem{Sreejith2019} G. J. Sreejith, S. Powell, and A. Nahum, {\it Emergent SO$(5)$ Symmetry at the Columnar Ordering Transition in the Classical Cubic Dimer Model}, Phys. Rev. Lett. {\bf 122}, 080601 (2019).
\bibitem{Li2017} Z.-X. Li, Y.-F. Jiang, S.-K. Jian, and H. Yao, {\it Fermion-induced quantum critical points}, Nat. Commun. {\bf 8}, 314 (2017).
\bibitem{Scherer2016} M. M. Scherer, and I. F. Herbut, {\it Gauge-field-assisted Kekul\'{e} quantum criticality}, Phys. Rev. B {\bf 94}, 205136 (2016).
\bibitem{Classen2017} L. Classen, I. F. Herbut, and M. M. Scherer, {\it Fluctuation-induced continuous transition and quantum criticality in Dirac semimetals}, Phys. Rev. B {\bf 96}, 115132 (2017).
\bibitem{Jian2017A} S.-K. Jian, and H. Yao, {\it Fermion-induced quantum critical points in two-dimensional Dirac semimetals}, Phys. Rev. B {\bf 96}, 195162 (2017).
\bibitem{Jian2017B} S.-K. Jian, and H. Yao, {\it Fermion-induced quantum critical points in three-dimensional Weyl semimetals}, Phys. Rev. B {\bf 96}, 155112 (2017).
\bibitem{Torres2018} E. Torres, L. Classen, I. F. Herbut, and M. M. Scherer, {\it Fermion-induced quantum criticality with two length scales in Dirac systems}, Phys. Rev. B {\bf 97}, 125137 (2018).
\bibitem{LiB2019} B.-H. Li, Z.-X. Li, and H. Yao, {\it Fermion-induced quantum critical point in Dirac semimetals: a sign-problem-free quantum Monte Carlo study}, Phys. Rev. B {\bf 101}, 085105 (2020).
\bibitem{Roy} B. Roy and V. Juri\v{c}i\'{c}, {\it Fermionic multicriticality near Kekul\'{e} valence-bond ordering on a honeycomb lattice}, Phys. Rev. B {\bf 99}, 241103 (2019).
\bibitem{Zhou2016} Z. Zhou, D. Wang, Z. Y. Meng, Y. Wang, and C. Wu, {\it Mott insulating states and quantum phase transitions of correlated SU(2N) Dirac fermions} Phys. Rev. B {\bf 93}, 245157 (2016).

\bibitem{Herbut2006} I. F. Herbut, {\it Interactions and Phase Transitions on Graphene's Honeycomb Lattice}, Phys. Rev. Lett. {\bf 97}, 146401 (2006).
\bibitem{Honerkamp2008} C. Honerkamp, {\it Density Waves and Cooper Pairing on the Honeycomb Lattice}, Phys. Rev. Lett. {\bf 100}, 146404 (2008).
\bibitem{Herbut2009} I. F. Herbut, V. Juri\v{c}i\'{c}, and B. Roy, {\it Theory of interacting electrons on the honeycomb lattice}, Phys. Rev. B {\bf 79}, 085116 (2009).
\bibitem{Strack2010} P. Strack, S. Takei, and W. Metzner, {\it Anomalous scaling of fermions and order parameter fluctuations at quantum criticality}, Phys. Rev. B {\bf 81}, 125103 (2010).
\bibitem{Yao2015} Z.-X. Li, Y.-F. Jiang, and H. Yao, {\it Solving the fermion sign problem in quantum Monte Carlo simulations by Majorana representation}, Phys. Rev. B {\bf 91}, 241117(R) (2015).
\bibitem{Ihrig2018} B. Ihrig, L. N. Mihaila, and M. M. Scherer, {\it Critical behavior of Dirac fermions from perturbative renormalization}, Phys. Rev. B {\bf 98}, 125109 (2018).
\bibitem{Mihaila2017} L. N. Mihaila,  N. Zerf, B. Ihrig, I. F. Herbut, and M. M. Scherer, {\it Gross-Neveu-Yukawa model at three loops and Ising critical behavior of Dirac systems}, Phys. Rev. B {\bf 96}, 165133 (2017).
\bibitem{Zerf2017} N. Zerf, L. N. Mihaila, P. Marquard, I. F. Herbut, and M. M. Scherer, {\it Four-loop critical exponents for the Gross-Neveu-Yukawa models}, Phys. Rev. D {\bf 96}, 096010 (2017).
\bibitem{Meng2019} C. Chen, X. Y. Xu, Z. Y. Meng, and M. Hohenadler, {\it Charge-Density-Wave Transitions of Dirac Fermions Coupled to Phonons}, Phys. Rev. Lett. {\bf 122}, 077601 (2019).
\bibitem{Lang2019} T. C. Lang and A. M. L\"{a}uchli, {\it Quantum Monte Carlo Simulation of the Chiral Heisenberg Gross-Neveu-Yukawa Phase Transition with a Single Dirac Cone}, Phys. Rev. Lett. {\bf 123}, 137602 (2019).
\bibitem{herbutlor} B. Roy, V. Juri\v{c}i\'{c}, and I. F. Herbut, {\it Emergent Lorentz symmetry near fermionic quantum critical points in two and three dimensions}, JHEP {\bf 04}, 018 (2016).
\bibitem{Mengzy} Y. Liu, W. Wang, K. Sun, and Z. Y. Meng, {\it Designer Monte Carlo simulation for the Gross-Neveu-Yukawa transition}, Phys. Rev. B {\bf 101}, 064308 (2020).
\bibitem{Geim2009RMP} A. H. Castro Neto, F. Guinea, N. M. R. Peres, K. S. Novoselov, and A. K. Geim, {\it The electronic properties of graphene}, Rev. Mod. Phys. {\bf 81}, 109-162 (2009).
\bibitem{Kane2010RMP} M. Z. Hasan and C. L. Kane, {\it Colloquium: Topological insulators}, Rev. Mod. Phys. {\bf 82}, 3045-3067 (2010).
\bibitem{Zhang2011RMP} X.-L. Qi and S.-C. Zhang, {\it Topological insulators and superconductors}, Rev. Mod. Phys. {\bf 83}, 1057-1110 (2011).
%
\bibitem{Wess1974} J. Wess, and B. Zumino, {\it Supergauge transformations in four dimensions}, Nucl. Phys. B {\bf 70}, 39-50 (1974).
\bibitem{Weinberg} S. Weinberg, {\it The Quantum Theory of Fields: Supersymmetry}, (Cambridge University Press, Cambridge, 2000).
\bibitem{ssLee2007} S.-S. Lee, {\it Emergence of supersymmetry at a critical point of a lattice model}, Phys. Rev. B {\bf 76}, 075103 (2007).
\bibitem{Vishwanath2014} T. Grover, D. N. Sheng, and A. Vishwanath, {\it Emergent space-time supersymmetry at the boundary of a topological phase}, Science {\bf 344}, 280–283 (2014).
\bibitem{Gross1974} D. J. Gross, and A. Neveu, {\it Dynamical symmetry breaking in asymptotically free field theories}, Phys. Rev. D {\bf 10}, 3235 (1974).
\bibitem{Rosenstein1993} B. Rosenstein, H.-L. Yu, and A. Kovner, {\it Critical exponents of new universality classes}, Phys. Lett. B {\bf 314}, 381-386 (1993).
\bibitem{Wetterich1993} C. Wetterich, {\it  Exact evolution equation for the effective potential}, Phys. Lett. B {\bf 301}, 90-94 (1993).
\bibitem{Berges2002} J. Berges, N. Tetradis, and C. Wetterich, {\it Non-perturbative renormalization flow in quantum field theory and statistical physics}, Phys. Rep. {\bf 363}, 223-386 (2002).
\bibitem{Janssen2014B} L. Janssen and I. F. Herbut, {\it Antiferromagnetic critical point on graphene's honeycomb lattice: A functional renormalization group approach}, {\it Phys. Rev. B} {\bf 89}, 205403 (2014).
\bibitem{Gracey2018} J. A. Gracey, {\it Large $N$ critical exponents for the chiral Heisenberg Gross-Neveu universality class}, Phys. Rev. D {\bf 97}, 105009 (2018).
\bibitem{Knorr2018} B. Knorr, {\it Critical chiral Heisenberg model with the functional renormalization group}, Phys. Rev. B {\bf 97}, 075129 (2018).
\bibitem{Litim2001} D. F. Litim, {\it Optimized renormalization group flows}, Phys. Rev. D {\bf 64}, 105007 (2001).
\bibitem{Wilson1974} K. G. Wilson and J. Kogut, {\it The renormalization group and the $\varepsilon$ expansion}, Phys. Rep {\bf 12}, 75 (1974).
\bibitem{Yin2018} S. Yin, S.-K. Jian, and H. Yao, {\it Chiral Tricritical Point: A New Universality Class in Dirac Systems}, Phys. Rev. Lett. {\bf 120}, 215702 (2018).
\bibitem{susyimprove} H. Gies, T. Hellwig, A. Wipf, and O. Zanusso, {\it A functional perspective on emergent supersymmetry}, Journal of High Energy Physics, {\bf 2017}, 132.
\bibitem{Aharony1997} O. Aharony, A. Hanany, K. Intriligator, N. Seiberg, and M. J. Strassler, {\it Aspects of $\mathcal{N}=2$ supersymmetric gauge theories in three dimensions}, Nucl. Phys. B {\bf 499}, 67 (1997).
\bibitem{Stephanov1995} M. A. Stephanov, {\it Dimensional reduction and quantum-to-classical reduction at high temperatures}, Phys. Rev. D {\bf 52}, 3746 (1995).
\bibitem{Hesselmann2016} S. Hesselmann and S. Wessel, {\it Thermal Ising transitions in the vicinity of two-dimensional quantum critical points}, Phys. Rev. B {\bf 93}, 155157 (2016).
\bibitem{Kato2015} Y. Kato and T. Misawa, {\it Quantum tricriticality in antiferromagnetic Ising model with transverse field: A quantum Monte Carlo study}, Phys. Rev. B {\bf 92}, 174419 (2015).
\bibitem{Jiansusy} S.-K. Jian, Y.-F. Jiang, and H. Yao, {\it Emergent Spacetime Supersymmetry in $3$D Weyl Semimetals and $2$D Dirac Semimetals}, Phys. Rev. Lett. {\bf 114}, 237001 (2015).
\bibitem{Lisusy2} Z.-X. Li, A. Vaezi, C. B. Mendl, and H. Yao, {\it Numerical observation of emergent spacetime supersymmetry at quantum criticality}, Sci. Adv. {\bf 4}, eaau1463 (2018).
\bibitem{Rahmanisusy} A. Rahmani, X. Zhu, M. Franz, and I. Affleck, {\it Emergent Supersymmetry from Strongly Interacting Majorana Zero Modes}, Phys. Rev. Lett. {\bf 115}, 166401 (2015).
\bibitem{Maciejkosusy} W. Witczak-Krempa and J. Maciejko, {\it Optical Conductivity of Topological Surface States with Emergent Supersymmetry}, Phys. Rev. Lett. {\bf 116}, 100402 (2016).
\bibitem{Greif2015} D. Greif, G. Jotzu, M. Messer, R. Desbuquois, and T. Esslinger, {\it Formation and Dynamics of Antiferromagnetic Correlations in Tunable Optical Lattices}, Phys. Rev. Lett. {\bf 115}, 260401 (2015).
\bibitem{Bloch2017} C. Gross and I. Bloch, {\it Quantum simulations with ultracold atoms in optical lattices}, Science {\bf 357}, 995-1001 (2017).




















\end{thebibliography}
\end{document}